\documentclass[sigconf,screen]{acmart}

\AtBeginDocument{%
  \providecommand\BibTeX{{%
    \normalfont B\kern-0.5em{\scshape i\kern-0.25em b}\kern-0.8em\TeX}}}

\usepackage{graphicx}
\usepackage{amsmath,amsfonts,amsthm}
\usepackage{array}
\usepackage{xcolor}

\usepackage{bbm}
\usepackage{bm}
\usepackage{multicol}
\usepackage{lipsum}
\usepackage{wrapfig}
\usepackage{arydshln}
\usepackage{subfig}
\usepackage{multirow}
\usepackage{wrapfig}
\usepackage[linesnumbered,ruled]{algorithm2e}
\usepackage{enumitem}

\newcommand\ddfrac[2]{{\displaystyle\frac{\displaystyle #1}{\displaystyle #2}}}

\newcommand{\diff}{\text{d}}

\newcommand{\Ical}{\mathcal{I}}

\newcommand{\Rbb}{\mathbb{R}}
\newcommand{\zbf}{\mathbf{z}}
\newcommand{\Rbf}{\mathbf{R}}

\newcommand{\phibf}{\boldsymbol \phi}
\newcommand{\thetabf}{\boldsymbol \theta}

\newcommand{\Xbf}{\mathbf{X}}

\newcommand{\inv}{^{\raisebox{.2ex}{$\scriptscriptstyle-1$}}}

\newtheorem{remark}{Remark}
\newtheorem{theorem}{Theorem}

\newtheorem{example}{Example}

\copyrightyear{2021}
\acmYear{2021}
\setcopyright{acmlicensed}
\acmConference[KDD '21]{Proceedings of the 27th ACM SIGKDD Conference on Knowledge Discovery and Data Mining}{August 14--18, 2021}{Virtual Event, Singapore}
\acmBooktitle{Proceedings of the 27th ACM SIGKDD Conference on Knowledge Discovery and Data Mining (KDD '21), August 14--18, 2021, Virtual Event, Singapore}
\acmPrice{15.00}
\acmDOI{10.1145/3447548.3467192}
\acmISBN{978-1-4503-8332-5/21/08}

\begin{document}

\title{Towards the D-Optimal Online Experiment Design for Recommender Selection}


\author{Da Xu}
\affiliation{%
  \institution{Walmart Labs}
  \city{Sunnyvale}
  \state{California}
  \country{USA}
}
\email{Da.Xu@walmartlabs.com}

\author{Chuanwei Ruan}
\authornote{The author is now with Instacart.}
\affiliation{%
  \institution{Walmart Labs}
  \city{Sunnyvale}
  \state{California}
  \country{USA}
}
\email{RuanChuanwei@gmail.com}

\author{Evren Korpeoglu}
\affiliation{%
  \institution{Walmart Labs}
  \city{Sunnyvale}
  \state{California}
  \country{USA}
}
\email{EKorpeoglu@walmart.com}

\author{Sushant Kumar}
\affiliation{%
  \institution{Walmart Labs}
  \city{Sunnyvale}
  \state{California}
  \country{USA}
}
\email{SKumar4@walmartlabs.com}

\author{Kannan Achan}
\affiliation{%
  \institution{Walmart Labs}
  \city{Sunnyvale}
  \state{California}
  \country{USA}
}
\email{KAchan@walmartlabs.com}


\begin{abstract}
Selecting the optimal recommender via online exploration-exploitation is catching increasing attention where the traditional A/B testing can be slow and costly, and offline evaluations are prone to the bias of history data. Finding the optimal online experiment is nontrivial since both the users and displayed recommendations carry contextual features that are informative to the reward. While the problem can be formalized via the lens of multi-armed bandits, the existing solutions are found less satisfactorily because the general methodologies do not account for the case-specific structures, particularly for the e-commerce recommendation we study. To fill in the gap, we leverage the \emph{D-optimal design} from the classical statistics literature to achieve the maximum information gain during exploration, and reveal how it fits seamlessly with the modern infrastructure of online inference. To demonstrate the effectiveness of the optimal designs, we provide semi-synthetic simulation studies with published code and data for reproducibility purposes. We then use our deployment example on Walmart.com to fully illustrate the practical insights and effectiveness of the proposed methods.
\end{abstract}



\maketitle

\section{Introduction}
\label{sec:introduction}
Developing and testing recommenders to optimize business goals are among the primary focuses of e-commerce machine learning. A crucial discrepancy between the business and machine learning world is that the target metrics, such as gross merchandise value (GMV), are difficult to interpret as tangible learning objectives. While a handful of surrogate losses and evaluation metrics have been found with particular empirical success \cite{shani2011evaluating,herlocker2004evaluating}, online experimentation is perhaps the only rule-of-thumb for testing a candidate recommender's real-world performance. In particular, there is a broad consensus on the various types of bias in the collected history data \cite{chen2020bias}, which can cause the "feedback-loop effect" if the empirical metrics are used without correction \cite{gilotte2018offline}. Recently, there has been a surge of innovations in refining online A/B testings and correcting offline evaluation methods \cite{yin2019identification,lee2018winner,xie2018false,gilotte2018offline}. However, they still fall short in specific applications where either the complexity of the problem outweighs their potential benefits, or their assumptions are not satisfied. Since our work discusses the e-commerce recommendations primarily, we assume the online shopping setting throughout the paper. On the A/B testing side, there is an increasing demand for interleaving the tested recommenders and targeted customers due to the growing interests in personalization \cite{schuth2015predicting}. The process of collecting enough observations and drawing inference with decent power is often slow and costly (in addition to the complication of defining the buckets in advance), since the number of combinations can grow exponentially. As for the recent advancements for offline A/B testing \cite{gilotte2018offline}, even though certain unbiasedness and optimality results have been shown in theory, the real-world performance still depends on the fundamental causal assumptions \cite{imbens2015causal}, e.g. unconfounding, overlapping and identifiability, which are rarely fulfilled in practice \cite{xu2020adversarial}. We point out that the role of both online and offline testings are irreplaceable regardless of their drawbacks; however, the current issues motivate us to discover more efficient solutions which can better leverage the randomized design of traffic.  

The production scenario that motivates our work is to choose from several candidate recommenders who have shown comparable performances in offline evaluation. By the segment analysis, we find each recommender more favorable to specific customer groups, but again the conclusion cannot be drawn entirely due to the exposure and selection bias in the history data. In other words, while it is safe to launch each candidate online, we still need randomized experiments to explore each candidates' real-world performance for different customer groups. We want to design the experiment by accounting for the customer features (e.g. their segmentation information) to minimize the cost of trying suboptimal recommenders on a customer group. Notice that our goal deviates from the traditional controlled experiments because we care more about minimizing the cost than drawing rigorous inference. In the sequel, we characterize our mission as a recommender-wise exploration-exploitation problem, a novel application to the best of our knowledge. Before we proceed, we illustrate the fundamental differences between our problem and learning the ensembles of recommenders \cite{jahrer2010combining}. The business metrics, such as GMV, are random quantities that depend on the recommended contents as well as the distributions that govern customers' decision-making. Even if we have access to those distributions, we never know in advance the conditional distribution given the recommended contents. Therefore, the problem can not be appropriately described by any fixed objective for learning the recommender ensembles.

In our case, the exploration-exploitation strategy can be viewed as a sequential game between the developer and the customers. In each round $t=1,\ldots,n$, where the role of $n$ will be made clear later, the developer chooses a recommender $a_t \in \{1,\ldots,k\}$ that produces the content $c_t$, e.g. top-k recommendations, according to the front-end request $r_t$, e.g. customer id, user features, page type, etc. Then the customer reveal the \textsl{reward} $y_t$ such as click or not-click. The problem setting resembles that of the \textsl{multi-armed bandits (MAB)} by viewing each recommender as the \emph{action} (\emph{arm}). The front-end request $r_t$, together with the recommended content $c_t = a_t(x_t)$, can be think of as the \emph{context}. Obviously, the context is informative of the reward because the clicking will depend on how well the content matches the request. On the other hand, an (randomized) experiment design can be characterized by a distribution $\pi$ over the candidate recommenders, i.e. $0 < \pi_t(a) < 1$, $\sum_{i=1}^k \pi_t(i)=1$ for $i=1,\ldots,n$. We point out that a formal difference between our setting and classical \emph{contextual bandit} is that the context here depends on the candidate actions. Nevertheless, its impact becomes negligible if choosing the best set of contents is still equivalent to choosing the optimal action. Consequently, the goal of finding the optimal experimentation can be readily converted to optimizing $\pi_t$, which is aligned with the bandit problems. The intuition is that by optimizing $\pi_t$, we refine the estimation of the structures between context and reward, e.g. via supervised learning, at a low exploration cost. 

The critical concern of doing exploration in e-commerce, perhaps more worrying than the other domains, is that irrelevant recommendations can severely harm user experience and stickiness, which directly relates to GMV. Therefore, it is essential to leverage the problem-specific information, both the contextual structures and prior knowledge, to further design the randomized strategy for higher efficiency. We use the following toy example to illustrate our argument.

\begin{example}
\label{example:one}
Suppose that there are six items in total, and the front-end request consists of a uni-variate user feature $r_t \in \Rbb$. The reward mechanism is given by the linear model:
\[ 
Y_t = \theta_1 \cdot I_1 \times X_t + \ldots + \theta_6 \cdot I_6 \times X_t;
\]
where $I_j$ is the indicator variable on whether item $j$ is recommended. Consider the top-3 recommendation from four candidate recommenders as follow (in the format of one-hot encoding): \\
\begin{equation}
\begin{split}
      a_1(r_t) = [1, 1, 1, 0, 0, 0]; \\
      a_2(r_t) = [0, 0, 0, 1, 1, 1]; \\ 
      a_3(r_t) = [0, 0, 1, 0, 1, 1]; \\
      a_4(r_t) = [0, 0, 0, 1, 1, 1]. \\
\end{split}
\end{equation}
If each recommender is explored with the probability, the role of $a_1$ is underrated since it is the only recommender that provides information about $\theta_1$ and $\theta_2$. Also, $a_2$ and $a_4$ give the same outputs, so their exploration probability should be discounted by half. Similarly, the information provided by $S_3$ can be recovered by $S_1$ and $S_4$ (or $S_2$) combined, so there is a linear dependency structure we may leverage.
\end{example}

The example is representative of the real-world scenario, where the one-hot encodings and user features may simply be replaced by the pre-trained embeddings. By far, we provide an intuitive understanding of the benefits from good online experiment designs. In Section \ref{sec:background}, we introduce the notations and the formal background of bandit problems. We then summarize the relevant literature in Section \ref{sec:related}. In Section \ref{sec:method}, we present our optimal design methods and describe the corresponding online infrastructure. Both the simulation studies and real-world deployment analysis are provided in Section \ref{sec:experiment}.   
We summarize the major contributions as follow.
\begin{itemize}
    \item We provide a novel setting for online recommender selection via the lens of exploration-exploitation.
    \item We present an optimal experiment approach and describe the infrastructure and implementation.
    \item We provide both open-source simulation studies and real-world deployment results to illustrate the efficiency of the approaches studied.
\end{itemize}

\section{Background}
\label{sec:background}
\begin{table}[hb]
    \centering
    \begin{tabular}{|m{1.6cm}|m{6cm}|}
    \hline
        $k$, $m$, $n$ & The number of candidate actions (recommenders); the number of top recommendations; the number of exploration-exploitation rounds. (may not known in advance).  \\ \hline
        $R_t$, $A_t$, $Y_t$ & The frond-end request, action selected by the developer, and the reward at round $t$. \\ \hline 
        $\vec{h}_t$, $\pi(\cdot \,|\, \cdot)$ & The history data collected until round $t$, and the randomized strategy (policy) that maps the contexts and history data to a probability measure on the action space. \\ \hline
        $\Ical$, $a_i(\cdot)$ & The whole set of items and the $i^{\text{th}}$ candidate recommender, with $a_i(\cdot) \in \Ical^{\otimes m}$. \\ \hline
        $\phibf\big(r_t, a_i(r_t) \big)$ & The (engineered or pre-trained) feature representation in $\Rbb^d$, specifically for the $t^{\text{th}}$-round front-end request and the output contents for the $i^{\text{th}}$ recommender. \\ \hline 
        
    \end{tabular}
    \caption{A summary of the notations. By tradition, we use uppercase letters to denote random variables, and the corresponding lowercase letters as observations.}
    \label{tab:notation}
\end{table}

We start by concluding our notations in Table \ref{tab:notation}. By convention, we use lower and upper-case letters to denote scalars and random variables, and bold-font lower and upper-case letters to denote vectors and matrices. We use $[k]$ as a shorthand for the set of: $\{1,2,\ldots,k\}$.
The randomized experiment strategy (policy) is a mapping from the collected data to the recommenders, and it should maximize the overall reward $\sum_{t=1}^n Y_t$.
The interactive process of the online recommender selection can be described as follow. 

1. The developer receives a front-end request $r_t\sim P_{\text{request}}$.

2. The developer computes the feature representations that combines the request and outputs from all candidate recommender: $\Xbf_t := \big\{\phibf\big(r_t, a_i(r_t) \big)\big\}_{i=1}^k$. 

3. The developer chooses a recommender $a_t$ according to the randomized experiment strategy $\pi(a | \Xbf_t, \vec{h}_t )$. 

4. The customer reveals the reward $y_t$.

In particular, the selected recommender $a_t$ depends on the request, candidate outputs, as well as the history data:
\[ 
a_t \sim \pi\Big(a \,\Big| \, r_t, \big\{\phibf\big(r_t, a_i(r_t) \big)\big\}_{i=1}^k, \vec{h}_t\Big),
\]
and the observation we collect at each round is given by:
\[
\Big(r_t, \big\{\phibf\big(r_t, a_i(r_t) \big)\big\}_{i=1}^k, a_t, y_t \Big).
\]

We point out that compared with other bandit applications, the restriction on computation complexity per round is critical for real-world production. This is because the online selection experiment is essentially an additional layer on top of the candidate recommender systems, so the service will be called by tens of thousands of front-end requests per second.
Consequently, the context-free exploration-exploitation methods, whose strategies focus on the cumulative rewards: $Q(a) = \sum_{j=1}^t y_j 1[a=j]$ and number of appearances: $N(a)=\sum_{j=1}^t 1[a=j]$ (assume up to round $t$) for $a=1,\ldots,k$, are quite computationally feasible, e.g.
\begin{itemize}[leftmargin=*]
    \item \textbf{$\epsilon$-greedy}: explores with probability $\epsilon$ under the uniform exploration policy $\pi(a) = 1/k$, and selects $\arg\max_{a\in[k]}Q(a)$ otherwise (for exploitation);
    \item \textbf{UCB}: selects $\arg\max_{a\in[k]}Q(a)+CI(a)$, where $CI(a)$ characterizes the confidence interval of the action-specific reward $Q(a)$, and is given by: $\sqrt{\big(\log 1/\delta\big)\big/N(a)}$ for some pre-determined $\delta$.
\end{itemize}
The more sophisticated \textbf{Thompson sampling} equips the sequential game with a Bayesian environment such that the developer:
\begin{itemize}[leftmargin=*]
    \item selects $\arg\max_{a\in[k]} \tilde{Q}(a)$, where $\tilde{Q}(a)$ is sampled from the posterior distribution $\text{Beta}(\alpha_t, \beta_t)$, and $\alpha_t$ and $\beta_t$ combines the prior knowledge of average reward and the actual observed rewards.
\end{itemize}
For Thompson sampling, it is clear that the front-end computations can be simplified to calculating the uni-variate indices ($Q$, $N$, $\alpha$, $\beta$). For MAB, taking account of the context often requires employing a parametric reward model: $y_a = f_{\thetabf}\big(\phibf(r, a(r))\big)$, so during exploration, we may also update the model parameters $\thetabf$ using the collected data. Suppose we have an \emph{optimization oracle} that returns $\hat{\thetabf}$ by fitting the empirical observations, then all the above algorithms can be converted to the context-aware setting, e.g.
\begin{itemize}[leftmargin=*]
    \item \textbf{epoch-greedy}: explores under $\pi(a) = 1/k$ for a epoch, and selects $\arg\max_{a\in[k]} \hat{y}_a := f_{\hat{\thetabf}}\big(\phibf(r, a(r))\big)$ otherwise;
    \item \textbf{LinUCB}: by assuming the reward model is linear, it selects $\arg\max_{a\in[k]} \hat{y}_a + CI(a)$ where $CI(a)$ characterizes the confidence of the linear model's estimation;
    \item \textbf{Thompson sampling}: samples $\hat{\thetabf}$ from the reward-model-specific posterior distribution of $\thetabf$, and selects $\arg\max_{a\in[k]} \hat{y}_a$.
\end{itemize}

We point out that the per-round model parameter update via the optimization oracle, which often involves expensive real-time computations, is impractical for most online services. Therefore, we adopt the stage-wise setting that divides exploration and exploitation (similar to epoch-greedy). The design of $\pi$ thus becomes very challenging since we may not have access to the most updated $\hat{\thetabf}$. Therefore, it is important to take advantage of the structure of $f_{\thetabf}(\cdot)$, which motivates us to connect our problem with the optimal design methods in the classical statistics literature.

\section{Related Work}
\label{sec:related}
We briefly discuss the existing bandit algorithms and explain their implications to our problem. Depending on how we perceive the environment, the solutions can be categorized into the frequentist and Bayesian setting. On the frequentist side, the reward model plays an important part in designing algorithms that connect to the more general \emph{expert advice} framework \cite{chu2011contextual}. The \textsl{EXP4} and its variants are known as the theoretically optimal algorithms for the expert advice framework if the environment is adversarial \cite{auer2002nonstochastic,mcmahan2009tighter,beygelzimer2011contextual}. However, customers often have a neutral attitude for recommendations, so it is unnecessary to assume adversarialness. In a neutral environment, the \textsl{LinUCB} algorithm and its variants have been shown highly effective \cite{auer2002nonstochastic,chu2011contextual}. In particular, when the contexts are viewed as i.i.d samples, several regret-optimal variants of LinUCB have been proposed \cite{dudik2011efficient,agarwal2014taming}. Nevertheless, those solutions all require real-time model updates (via the optimization oracle), and are thus impractical as we discussed earlier.

On the other hand, several suboptimal algorithms that follow the \textsl{explore-then-commit} framework can be made computationally feasible for large-scale applications \cite{robbins1952some}. The key idea is to divide exploration and exploitation into different stages, like the \textsl{epoch-greedy} and \textsl{phased exploration} algorithms \cite{rusmevichientong2010linearly,abbasi2009forced,langford2008epoch}. The model training and parameter updates only consume the back-end resources dedicated for exploitation, and the majority of front-end resources still take care of the model inference and exploration. Therefore, the stage-wise approach appeals to our online recommender selection problem, and it resolves certain infrastructural considerations that we explain later in Section \ref{sec:method}.

On the Bayesian side, the most widely-acknowledge algorithms belong to the \textsl{Thompson sampling}, which has a long history and fruitful theoretical results \cite{thompson1933likelihood,chapelle2011empirical,russo2017tutorial,russo2016information}. When applied to contextual bandit problems, the original Thompson sampling also requires per-round parameter update for the reward model \cite{chapelle2011empirical}. Nevertheless, the flexibility of the Bayesian setting allows converting Thompson sampling to the stage-wise setting as well.

In terms of real-world applications, online advertisement and news recommendation \cite{li2010contextual,agrawal2012analysis,li2012unbiased} are perhaps the two major domains where contextual bandits are investigated. Bandits have also been applied to our related problems such as item recommendation \cite{kawale2015efficient,li2016collaborative,zhao2013interactive} and recommender ensemble \cite{canamares2019multi,xu2022advances}. To the best of our knowledge, none of the previous work studies contextual bandit for the recommender selection. 

\section{Methodologies}
\label{sec:method}
As we discussed in the literature review, the stage-wise (phased) exploration and exploitation appeals to our problem because of their computation advantage and deployment flexibility. To apply the stage-wise exploration-exploitation to online recommender selection, we describe a general framework in Algorithm \ref{algo:stage-wise}.

\begin{algorithm}
\SetAlgoLined
\KwIn{Reward model $f_{\thetabf}(\cdot)$; the restart criteria; the initialized history data $\vec{h}_t$.}
\While{total rounds $\leq$ n}{
\If{restart criteria is satisfied}{
Reset $\vec{h}_t$\;
}
Play $n_1$ rounds of random exploration, for instance: $\pi(a \,|\, \cdot) = \frac{1}{k}$, and collect observation to $\vec{h}_t$\;
Find the optimal $\hat{\thetabf}$ based on $\vec{h}_t$ (e.. via empirical-risk minimization)\;
Play $n_2$ rounds of exploitation with: $a_t = \arg\max_{a}f_{\hat{\thetabf}}\big(r_t, a(r_t)\big)$ \;
}
 \caption{Stage-wise exploration and exploitation}
 \label{algo:stage-wise}
\end{algorithm}

The algorithm is deployment-friendly because Step 5 only involves front-end and cache operation, Step 6 is essentially a batch-wise training on the back-end, and Step 7 applies directly to the standard front-end inference. Hence, the algorithm requires little modification from the existing infrastructure that supports real-time mdoel inference.
Several additional advantages of the stage-wise algorithms include:
\begin{itemize}
    \item the number of of exploration and exploitation rounds, which decides the proportion of traffic for each task, can be adaptively adjusted by the resource availability and response time service level agreements;
    \item the non-stationary environment, which are often detected via the hypothesis testing methods as described in \cite{bubeck2012best,luo2018efficient,auer2016algorithm}, can be handled by setting the restart criteria accordingly.
\end{itemize}

\subsection{Optimal designs for exploration}
\label{sec:optimal-design}

This section is dedicated to improving the efficiency of exploration in Step 5. Throughout this paper, we emphasize the importance of leveraging the case-specific structures to minimize the number of exploration steps it may take to collect equal information for estimating $\thetabf$. Recall from Example \ref{example:one} that one particular structure is the relation among the recommended contents, whose role can be thought of as the design matrix in linear regression. Towards that end, our goal is aligned with the optimal design in the classical statistics literature \cite{o2003gentle}, since both tasks aim at optimizing how the design matrix is constructed. Following the previous buildup, the reward model has one of the following forms:
\begin{equation}
\label{eqn:reward-model}
    y_t = \left\{ 
    \begin{array}{ll}
    \thetabf^{\intercal} \phibf\big(r_t, a(r_t) \big), & \text{ linear model} \\
    f_{\thetabf}\big(\phibf\big(r_t, a(r_t) \big) \big), & \text{ for some nonlinear }f_{\thetabf}(\cdot),
    \end{array}
    \right.
\end{equation}
We start with the frequentist setting, i.e. $\thetabf$ do not admit a prior distribution. In each round $t$, we try to find a optimal design $\pi(\cdot\, |\, \cdot)$ such that the action sampled from $\pi$ leads to a maximum information for estimating $\thetabf$. For statistical estimators, the Fisher information is a key quantity for evaluating the amount of information in the observations. For the general $f_{\thetabf}$, the Fisher information under (\ref{eqn:reward-model}) is given by:
\begin{equation}
\label{eqn:fisher-general}
    M(\pi) = \sum_{a_i=1}^k \pi(a_i)\, \nabla_{\thetabf} f_{\thetabf}\big(\phibf\big(r_t, a_i(r_t) \big) \big) \cdot  \nabla_{\thetabf} f_{\thetabf}\big(\phibf\big(r_t, a_i(r_t) \big) \big)^{\intercal},
\end{equation}
where $\pi(a_i)$ is a shorthand for the designed policy. For the linear reward model, the Fisher information is simplified to:
\begin{equation}
\label{eqn:fisher-general}
    M(\pi) = \sum_{a_i=1}^k \pi(a_i)\, \phibf\big(r_t, a_i(r_t) \big) \cdot \phibf\big(r_t, a_i(r_t) \big)^{\intercal}.
\end{equation}
To understand the role Fisher information in evaluating the underlying \emph{uncertainty} of a model, according to the textbook derivations for linear regression, we have:
\begin{itemize}
    \item $\text{var}(\hat{\thetabf}) \propto M(\pi)\inv$;
    \item the prediction variance for $\phibf_i := \phibf\big(r_t, a_i(r_t) \big)$ is given by $\text{var}(\hat{y}_i) \propto \phibf_i M(\pi)\inv \phibf_i^{\intercal}$. 
\end{itemize}
Therefore, the goal of optimal online experiment design can be explained as minimizing the uncertainty in the reward model, either for parameter estimation or prediction. In statistics, a \textsl{D-optimal design} minimizes $\det|M(\pi)\inv|$ from the perspective of estimation variance , and the \textsl{G-optimal design} minimize $\max_{i \in \{1,\ldots,k\}} \phibf_i M(\pi)\inv \phibf_i^{\intercal}$ from the perspective of prediction variance. A celebrated result states the equivalence between D-optimal and G-optimal designs.
\begin{theorem}[Kiefer-Wolfowitz \cite{kiefer1960equivalence}]
\label{thm:kw}
For a optimal design $\pi^*$, the following statements are equivalent:
\begin{itemize}
    \item $\pi^* =\max_{\pi} \log \det\big|M(\pi)\big|$;
    \item $\pi^*$ is D-optimal;
    \item $\pi^*$ is G-optimal.
\end{itemize}
\end{theorem}
Theorem \ref{thm:kw} suggests that we use convex optimization to find the optimal design for both parameter estimation and prediction:
\[
\max_{\pi} \log \det\big|M(\pi)\big|\, \text{ s.t.} \, \sum_{a_i=1}^k\pi(a_i)=1.
\]
However, a drawback of the above formulation is that it does not involve the observations collected in the previous exploration rounds. Also, the optimization problem does not apply to the Bayesian setting if we wish to use Thompson sampling. Luckily, we find that optimal design for the Bayesian setting has a nice connection to the above problem, and it also leads to a straightforward solution that utilizes the history data as a prior for the optimal design. 

We still assume the linear reward setting, and the prior for $\thetabf$ is given by $\thetabf \sim N(0, \Rbf)$ where $\Rbf$ is the covariance matrix. Unlike in the frequentist setting, the Bayesian design focus on the design optimality in terms of certain utility function $U(\pi)$. A common choice is the expected gain in \emph{Shannon information}, or equivalently, the Kullback-Leibler divergence between the prior and posterior distribution of $\thetabf$. The intuition is that the larger the divergence, the more information there is in the observations. Let $y$ be the hypothetical rewards for $\phibf\big(r_t, a_1(r_t) \big),\ldots, \phibf\big(r_t, a_k(r_t) \big) $. Then the gain in Shannon information is given by:
\begin{equation}
\begin{split}
    U(\pi) & = \int \log p(\thetabf | y, \pi) p(y, \thetabf | \pi)\diff \thetabf \diff y \\ 
    & = C + \frac{1}{2}\log\det \big|M(\pi) + \Rbf\inv \big|,
\end{split}
\end{equation}
where $C$ is a constant. Therefore, maximizing $U(\pi)$ is equivalent to maximizing $\log\det \big|M(\pi) + \Rbf\inv \big|$.

Compared with the objective for the frequentist setting, there is now an additive $\Rbf$ term inside the determinant. Notice that $\Rbf$ is the convariance of the prior, so given the previous history data, we can simply plug in the empirical estimation of $\Rbf$. In particular, let $\vec{\phibf}_{t-1}$ be the collection of feature vectors from the previous exploration rounds: $\big[\phibf\big(x_1, a_1(x_1), \ldots, \phibf\big(x_{t-1}, a_{t-1}(x_{t-1}) \big) \big]$. Then $\Rbf$ is simply estimated by $\big( \vec{\phibf}_{t-1} \vec{\phibf}_{t-1}^{\intercal} \big)\inv$. Therefore, the objective for Bayesian optimal design, after integrating the prior from the history data, is given by:
\begin{equation}
\label{eqn:bayesian-obj}
    \text{maximize}_{\pi} \log\det \big|M(\pi) + \lambda \phibf_{t-1} \phibf_{t-1}^{\intercal} \big|,\, \text{s.t.}\, \sum_{a_i=1}^k \pi(a_i) = 1,
\end{equation}
where we introduce the hyper parameter $\lambda$ to control influence of the history data.
We refer the interested readers to \cite{covey1970optimal,dykstra1971augmentation,mayer1973method,johnson1983some} for the historic development of this topic.

Moving beyond the linear setting, the results stated in the Kiefer-Wolfowitz theorem also holds for nonlinear reward model \cite{white1973extension}. Unfortunately, it is very challenging to find the exact optimal design when the reward model is nonlinear \cite{yang2013optimal}. The difficulty lies in the fact that $\nabla_{\thetabf} f_{\thetabf}\big(\phibf\big(r_t, a(r_t) \big) \big)$ now depends on $\thetabf$, so the Fisher information (\ref{eqn:fisher-general}) is also a function of the unknown $\thetabf$. One solution which we find computationally feasible is to consider a local linearization of $f_{\thetabf}(\cdot)$ using the Taylor expansion:
\begin{equation}
\begin{split}
    f_{\thetabf}\big(\phibf\big(r_t, a(r_t)\big)\big) & \approx f_{\thetabf_0}\big(\phibf\big(r_t, a(r_t)\big)\big) + \\ 
    & \qquad \quad \nabla_{\thetabf} f_{\thetabf}\big(\phibf\big(r_t, a(r_t)\big)\big) \big | _{\thetabf=\thetabf_0} \big(\thetabf - \thetabf_0 \big),
\end{split}
\end{equation}
where $\thetabf_0$ is some local approximation. In this way, we have:
\begin{equation}
\label{eqn:grad-approx}
    \nabla_{\thetabf} f_{\thetabf} \approx \nabla_{\thetabf} f_{\thetabf}\big(\phibf\big(r_t, a(r_t)\big)\big) \big | _{\thetabf=\thetabf_0},
\end{equation}
and the local Fisher information will be given by $M(\pi;\thetabf_0)$ according to (\ref{eqn:fisher-general}). The effectiveness of local linearization completely depends on the choice of $\thetabf_0$. Using the data gathered from previous exploration rounds is a reasonable way to estimate $\thetabf_0$. We plug in $\thetabf_0$ to obtain the optimal designs for the following exploration rounds:
\begin{equation}
\label{eqn:nonlinear-obj}
\text{maximize}_{\pi}\log\det\big|M(\pi;\thetabf_0)\big|\, \text{ s.t.}\, \sum_{a_i=1}^k \pi(a_i) = 1 .
\end{equation}
We do not study the Bayesian optimal design under the nonlinear setting because even the linearization trick will be complicated. Moreover, by the way we construct $\thetabf_0$ above, we already pass a certain amount of prior information to the design. 

\begin{remark}
Before we proceed, we briefly discuss what to expect from the optimal design in theory. Optimizing the $\log\det |\cdot|$ objective is essentially finding the minimum-volume ellipsoid, also known as the \emph{John ellipsoid} \cite{hazan2016volumetric}. According to the previous results from the geometric studies, using the proposed optimal design will do no worse than the uniform exploration if the reward model is misspecified. Also, we can expect an average $\sqrt{d}$ improvement in the frequentist's linear reward setting \cite{bubeck2012towards}, which means it only takes $o(1/\sqrt{d})$ of the previous exploration steps to estimate $\thetabf$ to the same precision. 
\end{remark}


\subsection{Algorithms}
\label{sec:algo}

In this section, we introduce an efficient algorithm to solve the optimal designs in (\ref{eqn:nonlinear-obj}) and (\ref{eqn:bayesian-obj}). We then couple the optimal designs to the stage-wise exploration-exploitation algorithms. The infrastructure for our real-world production is also discussed. 
We have shown earlier that finding the optimal design requires solving a convex optimization programming. Since the problem is often of moderate size as we do not expect the number of recommenders $k$ to be large, we find the \textsl{Frank-Wolfe} algorithm highly efficient \cite{frank1956algorithm,jaggi2013revisiting}. We outline the solution for the most general non-linear reward case in Algorithm \ref{algo:solver}. The solutions for the other scenarios are included as special cases, e.g. by replacing $M(\pi)$ with $M(\pi) + \Rbf\inv$ for the Bayesian setting.

\begin{algorithm}
\SetAlgoLined
\KwIn{A subroutine for computing the $M(\pi;\thetabf_0)$ in (\ref{eqn:fisher-general}) or (\ref{eqn:grad-approx}); the estimation $\thetabf_0 = \hat{\thetabf}$ and $\eta_a := \nabla_{\thetabf} f_{\thetabf}\big(\phibf\big(r_t, a(r_t) \big)\big)\big |_{\thetabf=\thetabf_0}$; the convergence criteria.}
Initialize $\pi^{\text{old}}$: $\pi^{\text{old}}(a) = \frac{1}{k}$, $a=1,\ldots,k$\;
\While{convergence criteria not met}{
find $\tilde{a} =\arg\max_{a} \eta_a^{\intercal} M\big(\pi^{\text{old}};\thetabf_0\big)^{\inv} \eta_a$ \;
compute $\lambda_a = \ddfrac{\eta_{\tilde{a}}^{\intercal} M\big(\pi^{\text{old}};\thetabf_0\big)^{\inv}\eta_{\tilde{a}} \big/d - 1}{ \eta_a^{\intercal} M\big(\pi^{\text{old}};\thetabf_0\big)^{\inv}\eta_a - 1}$ \;
\For{$a=1,\ldots,k$}{
$\pi^{\text{new}}(a) = (1-\lambda_a)\pi^{\text{old}}(a) + \lambda_a 1 [a=\tilde{a}]$ \;
}
$\pi^{\text{old}} = \pi^{\text{new}}$
}
 \caption{The optimal design solver}
 \label{algo:solver}
\end{algorithm}
Referring to the standard analysis of Frank-Wolfe algorithm \cite{jaggi2013revisiting}, we show that the it takes the solver at most $\mathcal{O}(d\log\log k + d/\epsilon)$ updates to achieve a multiplicative $(1+\epsilon)$ optimal solution. Each update has an $\mathcal{O}(kd^3)$ computation complexity, but $d$ is usually small in practice (e.g. $d=6$ in Example \ref{example:one}), which we will illustrate with more detail in Section \ref{sec:experiment}.

By treating the optimal design solver as a subroutine, we now present the complete picture of the stage-wise exploration-exploitation with optimal design.
To avoid unnecessary repetitions, we describe the algorithms for nonlinear reward model under frequentist setting (Algorithm \ref{algo:batch-greedy}), and for linear reward model under the Thompson sampling. They include the other scenarios as special cases.
\begin{algorithm}
\SetAlgoLined
\KwIn{Reward model $f_{\thetabf}(\cdot)$; restart criteria; initialize history data $\vec{h}_t$; optimal design solver; initialize $\hat{\thetabf}$.}
\While{total rounds $\leq$ n}{
\If{restart criteria is satisfied}{
Reset $\vec{h}_t$\;
}
Compute the optimal exploration policy $\pi^*$ using the optimal design solver under $\thetabf_0 = \hat{\thetabf}$;
Play $n_1$ rounds of exploration under $\pi^*$ and collect observation to $\vec{h}_t$\;
Optimize and update $\hat{\thetabf}$ based on $\vec{h}_t$\;
Play $n_2$ rounds of exploitation with: $a_t = \arg\max_{a\in[k]}f_{\hat{\thetabf}}\big(r_t, a(r_t)\big)$ \;
}
 \caption{Stage-wise exploration-exploitation with optimal design.}
 \label{algo:batch-greedy}
\end{algorithm}

To adapt the optimal design to the Bayesian setting, we only need to make a few changes to the above algorithm:
\begin{itemize}
    \item the optimal design solver is now specified for solving (\ref{eqn:bayesian-obj});
    \item instead of optimizing $\hat{\thetabf}$ via the empirical-risk minimization, we update the posterior of $\thetabf$ using the history data $\vec{h}_t$;
    \item in each exploitation round, we execute Algorithm \ref{algo:ts} instead.
\begin{algorithm}
\SetAlgoLined
Compute the posterior distribution $P_{\thetabf}$ according to the prior and collected data \;
Sample $\hat{\thetabf}$ from $P_{\thetabf}$\;
Select $a_t = \arg\max_{a} \hat{\thetabf}^{\intercal}\phibf\big(r_t, a(r_t)\big)$ \;
Collect the observation to $\vec{h}_t$\;
 \caption{Optimal design for Thompson sampling at exploitation rounds.}
 \label{algo:ts}
\end{algorithm}
\end{itemize}

For Thompson sampling, the computation complexity of exploration is the same as Algorithm \ref{algo:batch-greedy}. On the other hand, even with a conjugate prior distribution, the Bayesian linear regression has an unfriendly complexity for the posterior computations. Nevertheless, under our stage-wise setup, the heavy lifting can be done at the back-end in a batch-wise fashion, so the delay will not be significant. In our simulation studies, we observe comparable performances from Algorithm \ref{algo:batch-greedy} and \ref{algo:ts}. Nevertheless, each algorithm may experience specific tradeoff in the stage-wise setting, and we leave it to the future work to characterize their behaviors rigorously. 


\subsection{Deployment Infrastructure}
\label{sec:infra}

In the ideal setting, the online recommender selection can be viewed as another service layer, which we refer to as the \emph{system bandit service}, on top of the model service infrastructure. 

\begin{figure}[hbt]
    \centering
    \includegraphics[width=0.7\linewidth]{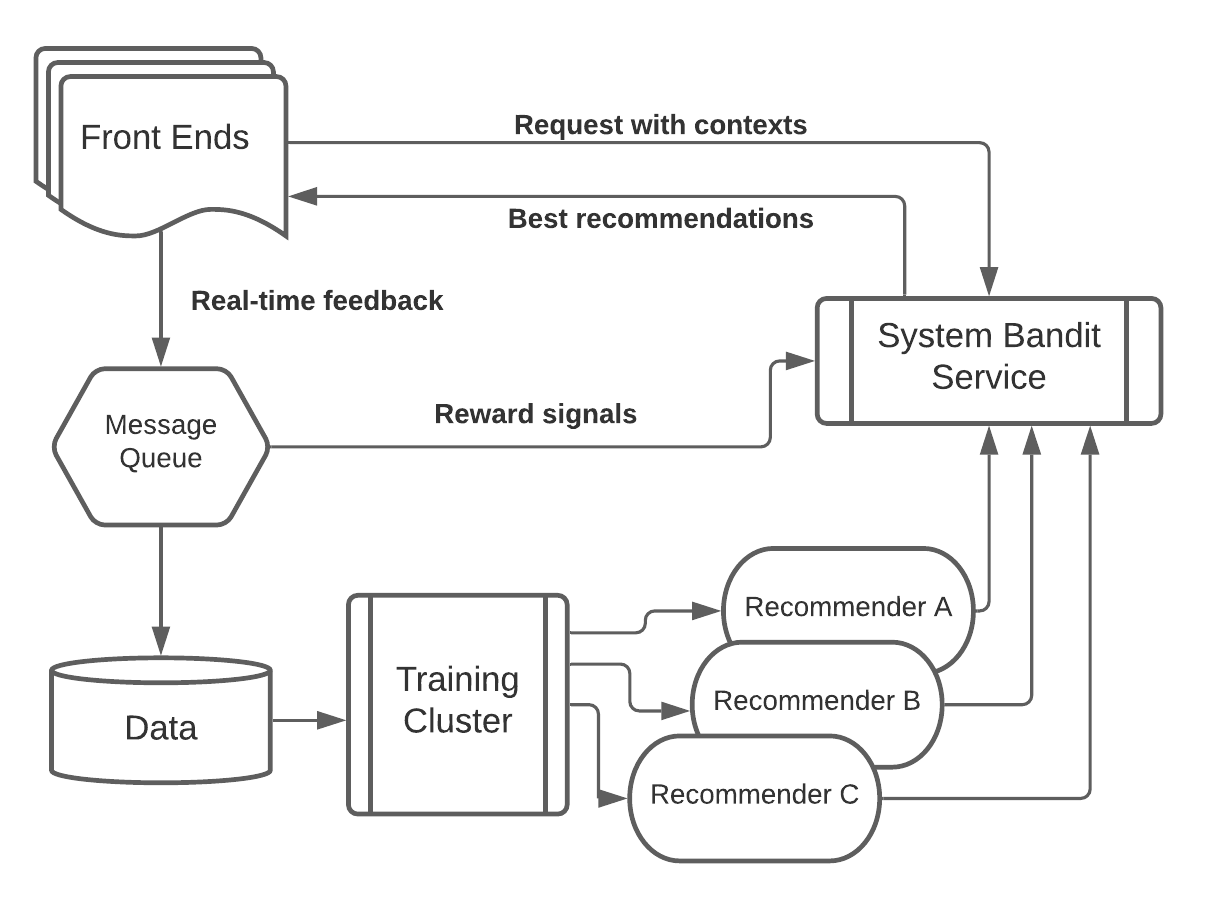}
    \caption{\small High-level overview of the system bandit service.}
    \label{fig:infra}
\end{figure}

An overview of the concept is provided in Figure \ref{fig:infra}. The system bandit module takes the request (which contains the relevant context), and the wrapped recommendation models. When the service is triggered, depending on the instruction from \emph{request distributor} (to explore or exploit), the module either queries the pre-computed reward, finds the best model and outputs its content, or run the optimal design solver using the pre-computed quantities and choose a recommender to explore. The request distributor is essential for keeping the resource availability and response time agreements, since the optimal-design computations can cause stress during the peak time. Also, we initiate the scoring (model inference) for the candidate recommenders in parallel to reduce the latency whenever there are spare resources. The pre-computations occur in the back-end training clusters, and their results (updated parameters, posterior of parameters, prior distributions) are stored in such as the mega cache for the front end. 

The logging system is another crucial component which maintains storage of the past reward signals, contexts, policy value, etc. The logging system works interactively with the training cluster to run the scheduled job for pre-computation. We mention that off-policy learning, which takes place at the back-end using the logged data, should be made robust to runtime uncertainties as suggested recently by \cite{xu2022towards}. Another detail is that the rewards are often not immediately available, e.g. for the conversion rate, so we set up an \emph{event stream} to collect the data. The system bandit service listens to the event streams and determines the rewards after each recommendation. 

\begin{figure}[hbt]
    \centering
    \includegraphics[width=0.9\linewidth]{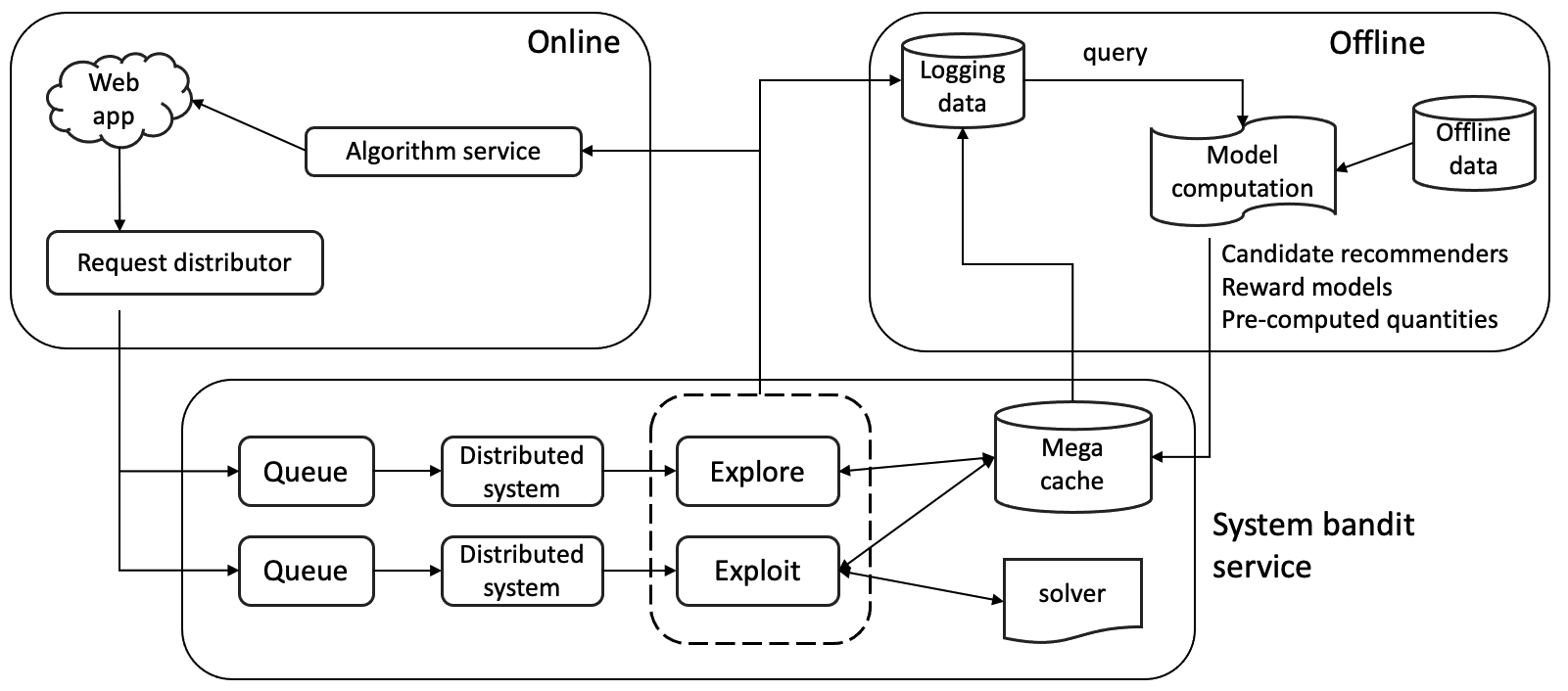}
    \caption{\small The deployment details of the optimal online experiment design for recommender selection.}
    \label{fig:deployment}
\end{figure}

For our deployment, we treat the system bandit serves as a middleware between the online and offline service. The details are presented in Figure \ref{fig:deployment}, where we put together the relevant components from the above discussion. It is not unusual these days to leverage the "near-line computation", and our approach takes the full advantage of the current infrastructure to support the optimal online experiment design for recommender selection. 

\section{Experiments}
\label{sec:experiment}
We first provide simulation studies to examine the effectiveness of the proposed optimal design approaches. We then discuss the relevant testing performance on \emph{Walmart.com}.

\subsection{Simulation}
For the illustration and reproducibility purposes, we implement the proposed online recommender selection under a semi-synthetic setting with a benchmark movie recommendation data. To fully reflect the exploration-exploitation dilemma in real-world production, we convert the benchmark dataset to an online setting such that it mimics the interactive process between the recommender and user behavior. A similar setting was also found in \cite{canamares2019multi} that studies the non-contextual bandits as model ensemble methods, with which we also compare in our experiments. We consider the linear reward model setting for our simulation\footnote{\url{https://github.com/StatsDLMathsRecomSys/D-optimal-recommender-selection}.}.

\textbf{Data-generating mechanism}. In the beginning stage, 10\% of the full data is selected as the training data to fit the candidate recommendation models, and the rest of the data is treated as the testing set which generates the interaction data adaptively. The procedure can be described as follow. In each epoch, we recommend one item to each user. If the item has received a non-zero rating from that particular user in the testing data, we move it to the training data and endow it with a positive label if the rating is high, e.g. $\geq 3$ under the five-point scale. Otherwise, we add the item to the rejection list and will not recommend it to this user again. After each epoch, we retrain the candidate models with both the past and the newly collected data. Similar to \cite{canamares2019multi}, we also use the \textbf{cumulative recall} as the performance metric, which is the ratio of the total number of successful recommendations (up to the current epoch) againt the total number of positive rating in the testing data. The reported results are averaged over ten runs.

\textbf{Dataset}. We use the \textit{MoiveLens 1M} \footnote{https://grouplens.org/datasets/movielens/1m/} dataset which consists of the ratings from 6,040 users for 3,706 movies. Each user rates the movies from zero to five. The movie ratings are binarized to $\{0,1\}$, i.e. $\geq 2.5$ or $<2.5$, and we use the metadata of movies and users as the contextual information for the reward model. In particular, we perform the one-hot transformation for the categorical data to obtain the feature mappings $\phibf(\cdot)$. For text features such as movie title, we train a word embedding model \cite{mikolov2013distributed} with 50 dimensions. The final representation is obtained by concatenating all the one-hot encoding and embedding. 

\begin{figure}[htb]
    \centering
    \includegraphics[width=\linewidth]{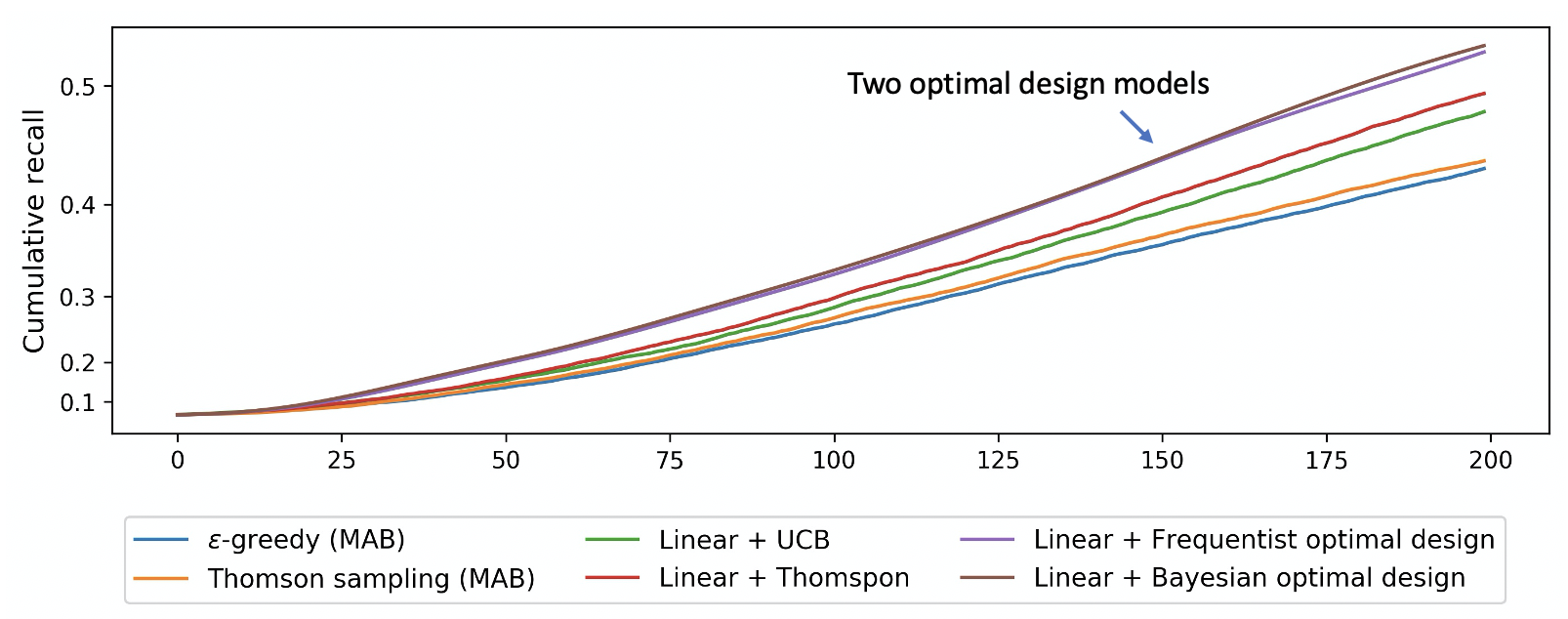}
    \caption{\small The comparisons of the cumulative recall (reward) per epoch for different bandit algorithms.}
    \label{fig:cumu_recall_abs}
\end{figure}

\textbf{Candidate recommenders}. We employ the four classical recommendation models: user-based collaborative filtering (CF) \cite{user_cf}, item-based CF \cite{item_cf}, popularity-based recommendation, and a matrix factorization model \cite{mat_fact}. To train the candidate recommenders during our simulations, we further split the 10\% initial training data into equal-sized training and validation dataset, for grid-searching the best hyperparameters. The validation is conducted by running the same generation mechanism for 20 epochs, and examine the performance for the last epoch.
For the user-based collaborative filtering, we set the number of nearest neighbors as 30. For item-based collaborative filtering, we compute the cosine similarity using the vector representations of movies. For relative item popularity model, the ranking is determined by the popularity of movies compared with the most-rated movies. For matrix factorization model, we adopt the same setting from \cite{canamares2019multi}. 

\textbf{Baselines}. To elaborate the performance of the proposed methods, we employ the widely-acknowledged exploration-exploitation algorithms as the baselines:
\begin{itemize}
    \item The multi-armed bandit (MAB) algorithm without context: \textbf{$\epsilon$-greedy} and \textbf{Thompson sampling}.
    \item Contextual bandit with the exploration conducted in the LinUCB fashion (\textbf{Linear+UCB}) and Thompson sampling fashion (\textbf{Linear+Thompson}).
\end{itemize}
We denote our algorithms by the \textbf{Linear+Frequentis optimal design} and the \textbf{Linear+Bayesian optimal design}.

\textbf{Ablation studies}\\
We conduct ablation studies with respect to the contexts and the optimal design component to show the effectiveness of the proposed algorithms. Firstly, we experiment on removing the user context information. Secondly, we experiment with our algorithm without using the optimal designs.

\textbf{Results}. The results on cumulative recall per epoch are provided in Figure \ref{fig:cumu_recall_abs}. It is evident that as the proposed algorithm with optimal design outperforms the other bandit algorithms by significant margins. In general, even though $\epsilon$-greedy gives the worst performance, the fact that it is improving over the epochs suggests the validity of our simulation setup. The Thompson sampling under MAB performs better than $\epsilon$-greedy, which is expected. The usefulness of context in the simulation is suggested by the slightly better performances from Linear+UCB and Linear+Thompson. However, they are outperformed by our proposed methods by significant margins, which suggests the advantage of leveraging the optimal design in the exploration phase. Finally, we observe that among the optimal design methods, the Bayesian setting gives a slightly better performance, which may suggest the usefulness of the extra steps in Algorithm \ref{algo:ts}. 

\begin{figure}[hbt]
    \centering
    \includegraphics[width=\linewidth]{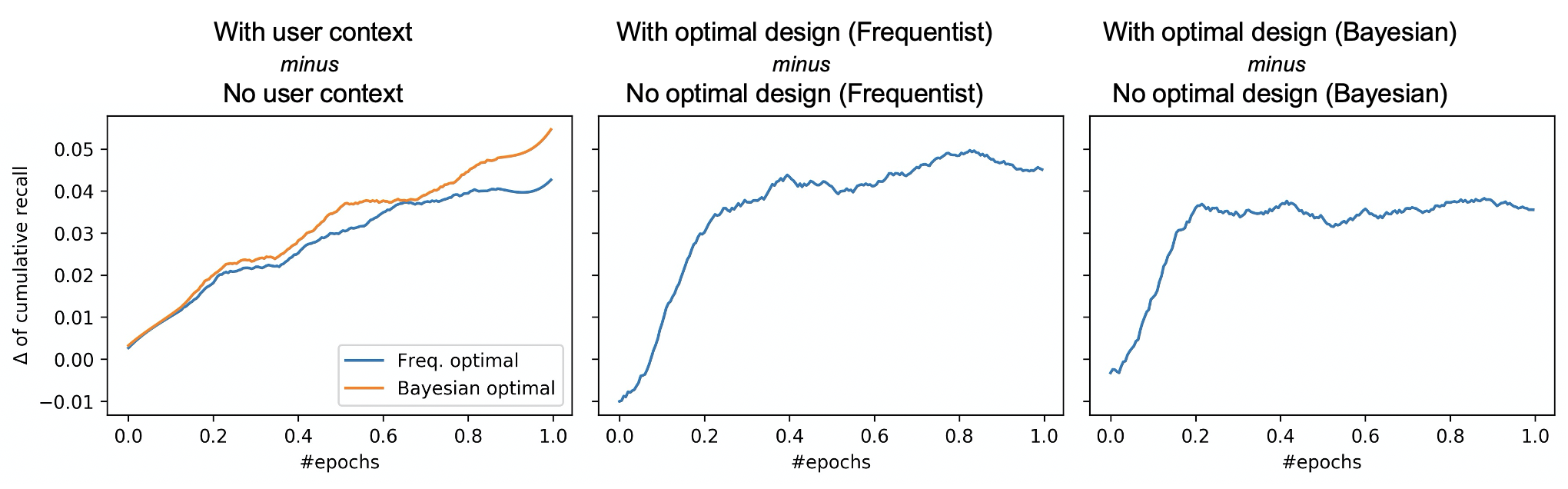}
    \caption{\small The difference in cumulative recall for system bandit under different settings. The left figure shows difference in performance between using and not using user context, under the frequentist and Bayesian setting, respectively. The other two figures compare using and not using the optimal design (uniform selection) in the exploration stages, both using the linear reward model. }
    \label{fig:cumu_recall_delta}
\end{figure}

The results for the ablation studies are provided in Figure \ref{fig:cumu_recall_delta}. The left-most plot shows the improvements from including contexts for bandit algorithms, and suggests that our approaches are indeed capturing and leveraging the signals of the user context. In the middle and right-most plots, we observe the clear advantage of conducting the optimal design, specially in the beginning phases of exploration, as the methods with optimal design outperforms their counterparts. We conjecture that this is because the optimal designs aim at maximizing the information for the limited options, which is more helpful when the majority of options have not been explored such as in the beginning stage of the simulation.

\begin{figure}[htb]
    \centering
    \includegraphics[width=0.8\linewidth]{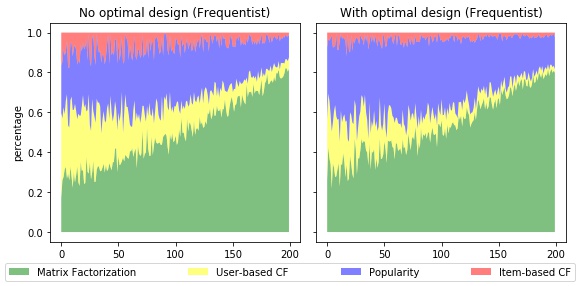}
    \caption{\small The percentage of traffic routed to each candidate recommender with and without using the optimal design under the frequentist setting. 
    }
    \label{fig:simu_arm_dist}
\end{figure}

Finally, we present a case study to fully illustrate the effectiveness of the optimal design, which is shown in Figure \ref{fig:simu_arm_dist}. It appears that in our simulation studies, the matrix factorization and popularity-based recommendation are found to be more effective. With the optimal design, the traffic concentrates more quickly to the two promising candidate recommenders than without the optimal design. The observations are in accordance with our previous conjecture that optimal design gives the algorithms more advantage in the beginning phases of explorations.



\subsection{Deployment analysis}
\label{sec:deployment}

\begin{wrapfigure}{r}{0.5\linewidth}
    \centering
    \includegraphics[width=\linewidth]{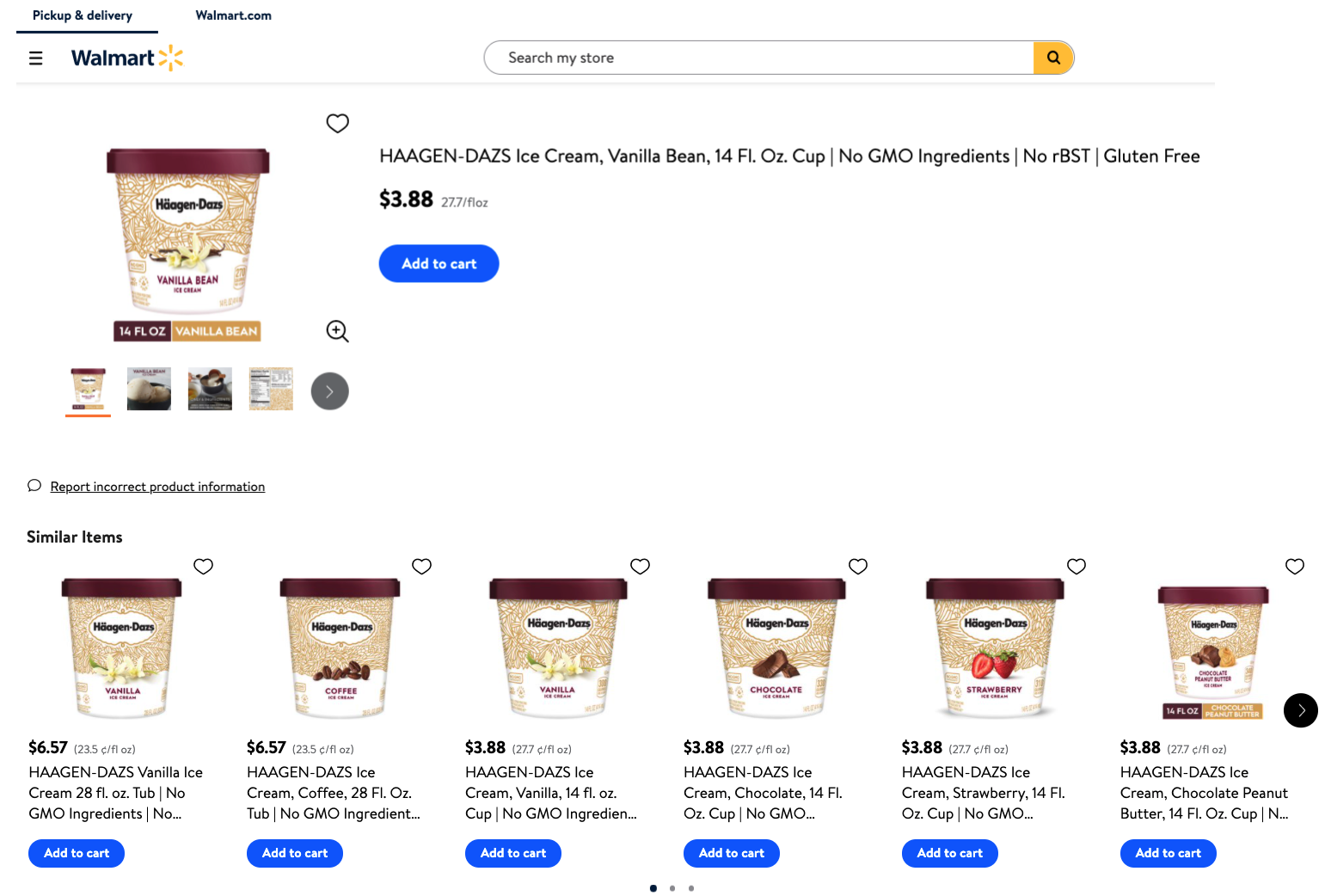}
    \caption{Item-page similar-item recommendation for groceries on Walmart.com.}
    \label{fig:example}
\end{wrapfigure}
We deployed our online recommender selection with optimal design to the similar-item recommendation of grocery items on Walmart.com. A webpage snapshot is provided in Figure \ref{fig:example}, where the recommendation appears on the item pages.
The baseline model for the similar-item recommendation is described in our previous work of \cite{xu2020methods}, and we experiment with three enhanced models that adjust the original recommendations based on the \emph{brand affinity}, \emph{price affinity} and \emph{flavor affinity}. We omit the details of each enhanced model since they are less relevant. The reward model leverages the item and user representations also described in our previous work. Specifically, the item embeddings are obtained from the Product Knowledge Graph embedding \cite{xu2020product}, and the user embeddings are constructed via the temporal user-item graph embedding \cite{xu2020inductive}. We adopt the frequentist setting where the reward is linear function of $\langle \text{item emb}, \text{user emb} \rangle$, plus some user and item contextual features:
\[ 
\theta_0 + \theta_1 \langle \zbf_{u}, \zbf_{I_1} \rangle + \ldots + \theta_m \langle \zbf_{u}, \zbf_{I_m} \rangle + \thetabf^{\intercal}[\text{user feats, items feats}],
\]
and $\zbf_u$ and $\zbf_I$ are the user and item embeddings. 

\begin{figure}[hbt]
    \centering
    \includegraphics[width=0.7\linewidth]{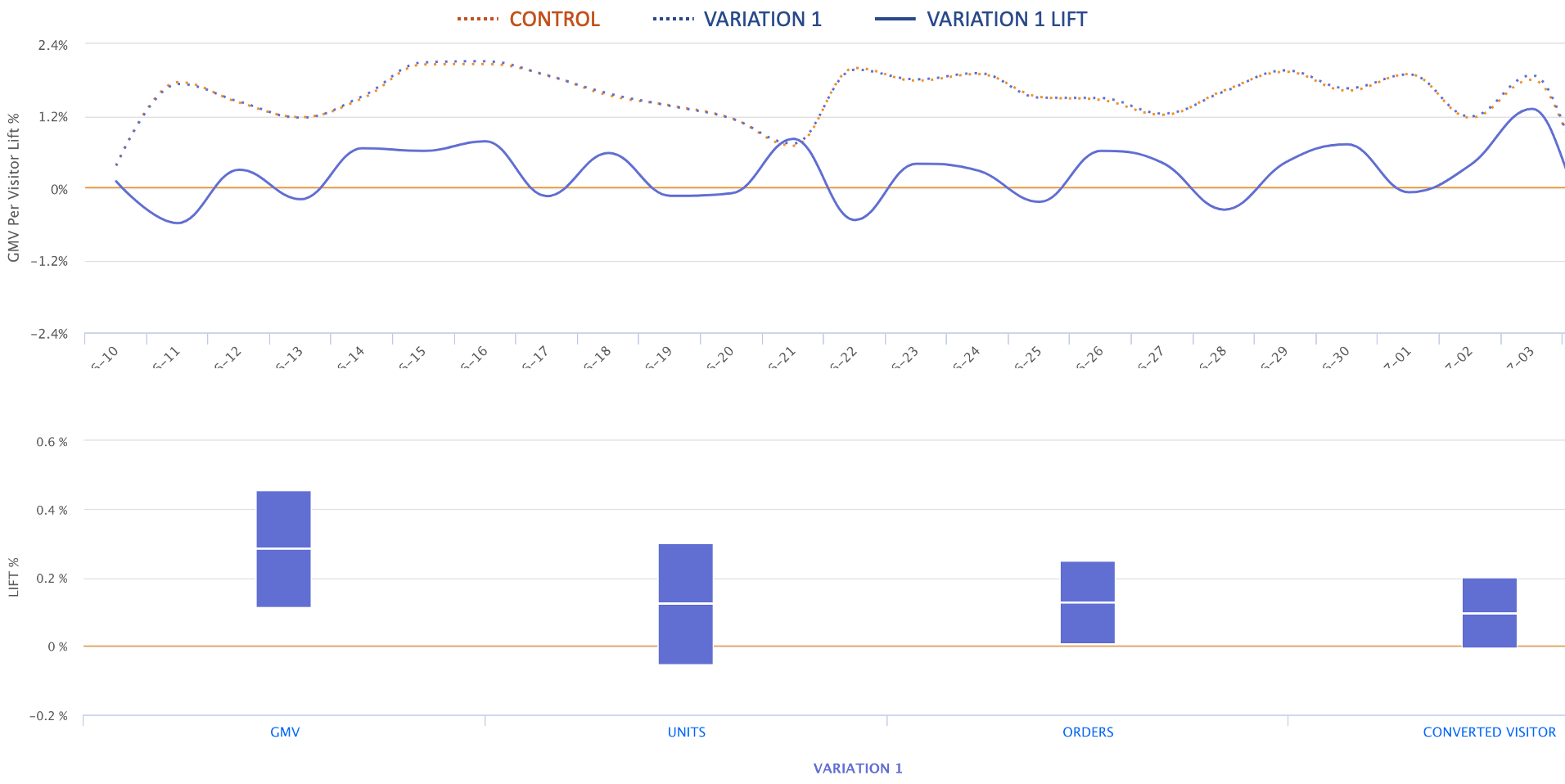}
    \caption{The testing results for the proposed stage-wise exploration-exploitation with optimal design.}
    \label{fig:ab-test}
\end{figure}

\begin{figure}[htb]
    \centering
    \includegraphics[width=0.8\linewidth]{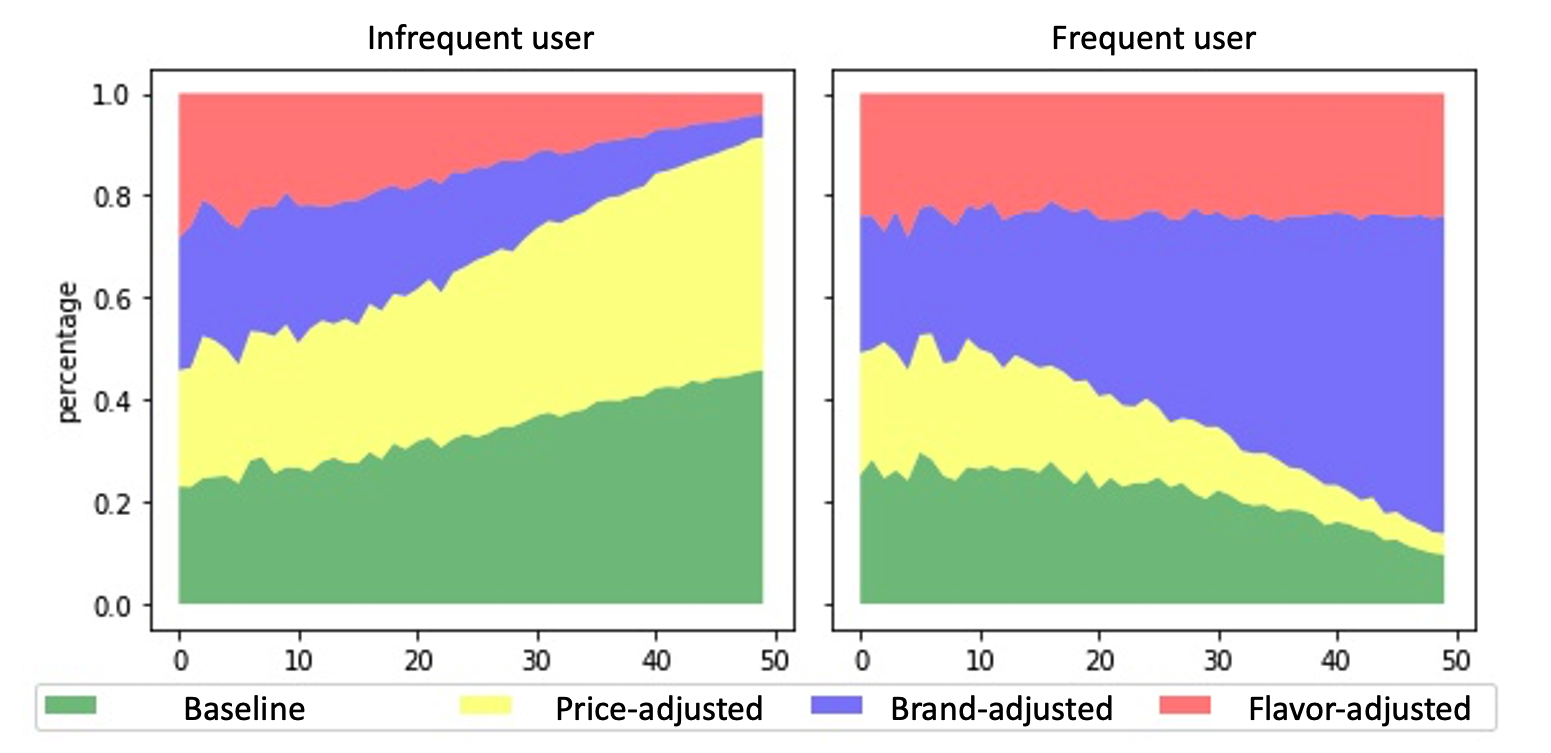}
    \caption{The analysis on the proportion of traffic directed to the frequent and infrequent customer. }
    \label{fig:prod-analysis}
\end{figure}

We conduct a posthoc analysis by examining the proportion of traffic directed to the frequent and infrequent user groups by the online recommender selection system (Figure \ref{fig:prod-analysis}). Interestingly, we observe different patterns where the brand and flavor-adjusted models serve the frequent customers more often, and the unadjusted baseline and price-adjusted model get more appearances for the infrequent customers. The results indicate that our online selection approach is actively exploring and exploiting the user-item features that eventually benefits the online performance. The simulation studies, on the other hand, reveal the superiority over the standard exploration-exploitation methods.

\section{Discussion}
\label{sec:discuss}

We study optimal experiment design for the critical online recommender selection. We propose a practical solution that optimizes the standard exploration-exploitation design and shows its effectiveness using simulation and real-world deployment results. 

\bibliographystyle{ACM-Reference-Format}
\bibliography{references}


\end{document}